\documentclass[prl,showpacs,twocolumn]{revtex4}
\usepackage{amsfonts}
\usepackage{mathrsfs}
\usepackage{amssymb}
\usepackage{pifont}
\usepackage{epsfig,subfigure,dsfont,amsthm,amsbsy,mathrsfs,amscd}

\def\be{\begin{equation}}
\def\ee{\end{equation}}
\def\ba{\begin{array}}
\def\ea{\end{array}}

\def\qed{\leavevmode\unskip\penalty9999 \hbox{}\nobreak\hfill
     \quad\hbox{\leavevmode  \hbox to.77778em{               \hfil\vrule   \vbox to.675em               {\hrule width.6em\vfil\hrule}\vrule\hfil}}
     \par\vskip3pt}

\input amssym.def

\begin{document}

\title{{\large \textbf{Sharing Bell nonlocality of bipartite high-dimensional pure states using only projective measurements}}}
\author{ Tinggui Zhang$^{1, \dag}$ Hong Yang$^{2}$ and Shao-Ming Fei$^{3}$}
\affiliation{${1}$ School of Mathematics and Statistics, Hainan Normal University,
Haikou, 571158, China \\
$2$ College of Physics and Electronic Engineering, Hainan Normal University,
Haikou, 571158, China\\
$3$ School of Mathematical Sciences, Capital Normal University, Beijing
100048, China \\
$^{\dag}$ Correspondence to tinggui333@163.com}

\begin{abstract}
Bell nonlocality is the key quantum resource in some device-independent
quantum information processing. It is of great importance to study the
efficient sharing of this resource. Unsharp measurements are widely used in
sharing the nonlocality of an entangled state shared among several
sequential observers. Recently, the authors in [Phys. Rev. Lett.$\mathbf{129}%
,230402(2022)$] showed that the Bell nonlocality of two-qubit pure states
can be shared even when one only uses projective measurements and local
randomness. We demonstrate that projective measurements are also sufficient
for sharing the Bell nonlocality of arbitrary high-dimensional pure
bipartite states. Our results promote further understanding of the
nonlocality sharing of high-dimensional quantum states under projective
measurements.
\end{abstract}

\maketitle

\bigskip

\section{Introduction}

Bell nonlocality, revealed by violating the Bell inequalities of quantum
entangled states, is one of the most startling predictions of quantum
mechanics \cite{ndsv}. It plays an important role in device independent
quantum information processing such as quantum key distribution \cite%
{jlak,wyym}, quantum secure direct communication \cite{zyjy,ylgl} and
communication complexity reduction \cite{hrsr,jgsf}.

In recent years, an interesting question about the shareability of Bell
nonlocality has been extensively studied \cite%
{rsng,smam,mlmg,mzxc,assd,dass,ctdh,akak,sdsd,glam,tcym,pjbr,ztfs,gmmh,smak,scll,cslt,cxwt,ztqx,ssap,jylq,mjhf,asat,yyxs,ssak,yzab,ztns,ymlz}%
. The question is whether the post-measurement state in a Bell experiment
can be re-used for showcasing nonlocality between several observers who
perform sequential quantum measurements, see FIG.1 (a) for the schematic
diagram. In 2015, Silva et al. showed that the Bell nonlocality from an
entangled pair can be utilized for multiple parties with sequential unsharp
measurements of intermediate strength \cite{rsng}. Since then, most of the
studies on nonlocality sharing adopt weak measurements or unsharp
measurements. In 2020, Brown et al. used average probability positive
operator-valued measure (POVM) \cite{pjbr} and showed that arbitrarily many
independent Bobs can share the nonlocality of the maximally entangled pure
two-qubit state $|\phi\rangle=\frac{1}{\sqrt{2}}(|00\rangle+|11\rangle)$
with the single Alice.

Various measurements have been used in demonstrating the Bell nonlocality.
Among them the projective measurement is the simplest one. Nevertheless, the
projective measurement is also the most destructive one to the quantum
states. Generally, an entangled state would become separable after such
measurements. Recently, in \cite{asat} the authors showed that if the Bobs
apply standard projective measurements (a random combination of three
projective measurement strategies with different probabilities, see FIG.1
(b)), then two and three Bobs can share the nonlocality of the two-qubit
state $|\phi\rangle$ with the single Alice. 
\begin{figure}[ptb]
\includegraphics[width=0.5\textwidth]{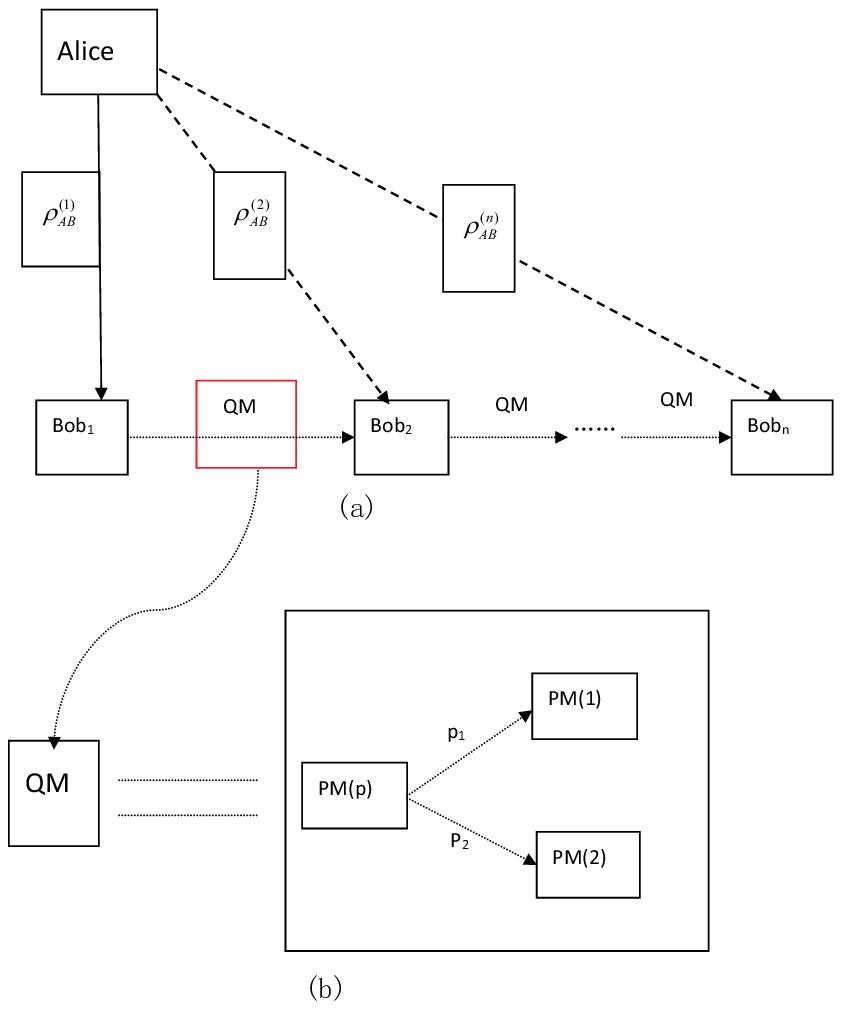}
\caption{(a) A quantum state $\protect\rho_{AB}^{(1)}$ is initially shared
by Alice and Bob$_{(1)}$. Bob$_{(1)}$ performs some kind of quantum
measurements on his part and then passes it to Bob$_{(2)}$. The
post-measurement state is $\protect\rho_{AB}^{(2)}$. Bob$_{(2)}$ measures $%
\protect\rho_{AB}^{(2)}$ on his part and passes it to Bob$_{(3)}$ and so on.
(b) Quantum measurement (QM) given by a random combination of several
projective measurements (PMs) with different probabilities $p$. Before the
experiment begins, all the parties agree to share correlated strings of
classical data $p=\{p_i\}$ satisfying $\sum_ip_i=1$.}
\end{figure}

The high-dimensional quantum systems can carry more information and are more
resistant to noises. High-dimensional quantum systems are important in
improving the performance of quantum networks, quantum key distribution,
quantum teleportation and quantum internet \cite{mkue,sdrh,mmaz,hkrb}.
Therefore, in this article we study the nonlocal correlation sharing
scenario for arbitrary high dimensional bipartite entangled pure states
along the line of \cite{asat} (Fig. 1). We show that projective measurement
is also a sufficient condition for two observers to share the Bell
nonlocality of any arbitrary dimensional bipartite entangled pure states.

\section{Nonlocal sharing of bipartite high-dimensional pure state}

Let $H_A$ and $H_B$ be Hilbert spaces with dimensions $dim(H_A)=s$ and $%
dim(H_B)=t$, respectively (without loss of generality, we assume $s \leq t$%
). A bipartite pure state $|\psi\rangle\in H_A\otimes H_B$ has the Schmidt
decomposition form, $|\psi\rangle=\sum_{i=1}^sc_i|i_A\rangle|i_B\rangle$,
where $c_i\in [0,1]$, $\sum_{i=1}^s c_i^2=1$, $\{i_A\}_1^s$ and $\{i_B\}_1^t$
are the orthonormal bases of $H_A$ and $H_B$, respectively. $|\psi\rangle$
is entangled if and only if at least two $c_i$s are nonzero. Without loss of
generality, below we assume that $c_i$ are arranged in descending order.

We focus on the sequential scenario shown in Fig. 1. To begin with, Alice
and Bob$_{1}$ share an arbitrary entangled bipartite pure state $%
\rho_{AB}^{(1)}=|\psi\rangle\langle\psi|$. Bob$_{k}$ $(k=1,2,\cdots,n)$ are
restricted to perform two different projective measurement settings: PM(1)
both choose projection measurement ($\lambda=1$)and PM(2) one chooses
projection measurement, the other chooses identity operator ($\lambda=2$).
Denote the binary input and output of Alice (Bob$_k$) by $X$ $%
(Y^{k})\in\{0,1\}$ and $A$ $(B^{(k)})\in\{0,1\}$, respectively. Before the
experiment begins, all the parties agree to share the correlated strings of
classical data $\lambda$ subjected to probability distribution $%
\{p_{\lambda}\}_{\lambda=1,2}$. Suppose Bob$_{1}$ performs the measurement
according to $Y^{1}=y$ with outcome $B^{1}=b$. Averaged over the inputs and
outputs of Bob$_{1}$, the unnormalized state shared between Alice and Bob$%
_{2}$ is given by 
\begin{equation}  \label{z001}
\rho_{AB}^{(2,\lambda)}=\frac{1}{2}\Sigma_{b,y}(I_s\otimes\sqrt{%
B_{b|y}^{(1,\lambda)}}) \rho_{AB}^{(1)}(I_s\otimes\sqrt{B_{b|y}^{(1,\lambda)}%
}),
\end{equation}
where $B^{(1,\lambda)}_{b|y}$ $(\lambda=1,2$) is the projective measurement (%
$(B^{(1,\lambda)}_{b|y})^2=B^{(1,\lambda)}_{b|y}$) corresponding to outcome $%
b$ of Bob$^{(1,\lambda)}$'s measurement for input $y$, $I_s$ is the $s\times
s$ identity matrix. Repeating this process, one gets the state $%
\rho_{AB}^{(k,\lambda)}$ shared between Alice and Bob$^{(k)}$, $%
k=2,3,\cdots,n$.

The Bell nonlocality is verified by the violation of the CHSH inequality 
\cite{jmar}. Each pair Alice-Bob$_{k}$ tests the CHSH inequality, 
\begin{equation}  \label{z002}
S_k\equiv\sum_{\lambda=1}^2 p_{\lambda}S_k^{\lambda}\leq 2,
\end{equation}
where 
\begin{equation}  \label{z003}
S_k^{\lambda}\equiv\sum_{x,y}(-1)^{xy}Tr(A_x\otimes
B_{y}^{(k,\lambda)})\rho_{AB}^{(k,\lambda)}.
\end{equation}
Here, $\{A_x,B_{y}^{(k,\lambda)}\}_{k=1,2,\cdots}$ denote the observables of
respective parties conditioned on $\lambda$. Only when $k=1$, $%
\rho_{AB}^{(1,\lambda)}=\rho_{AB}^{(1)}$.

Let us consider the simplest scenario, namely, $n=2$ and $s,t$ are even. We
set the Alice's quantum measurements to be given by observables 
\begin{equation}  \label{z004}
A_{0}=I_{\frac{s}{2}} \otimes(\cos\theta\sigma_3+\sin\theta\sigma_1),
\end{equation}
and 
\begin{equation}  \label{z005}
A_{1}=I_{\frac{s}{2}} \otimes(\cos\theta\sigma_3-\sin\theta\sigma_1)
\end{equation}
for some $\theta\in [0,\frac{\pi}{2}]$.

Case (i): ($\lambda =1$). We set the Bob$_{1}$'s projective measurements to
be given by observables 
\begin{equation}
B_{0|0}^{(1,1)}=\frac{1}{2}[I_{t}+(I_{\frac{t}{2}}\otimes \sigma _{3})]
\label{z006}
\end{equation}%
and 
\begin{equation}
B_{0|1}^{(1,1)}=\frac{1}{2}[I_{t}+(I_{\frac{t}{2}}\otimes \sigma _{1})].
\label{z007}
\end{equation}%
Denote $B_{1|y}^{(1,1)}=I_{t}-B_{0|y}^{(1,1)}$ and $%
B_{y}^{(1,1)}=B_{0|y}^{(1,1)}-B_{1|y}^{(1,1)}$ for $y=0,1$. Similar to the
calculations in Ref. \cite{ztfs}, it is not difficult to obtain that $%
S_{1}^{1}\geq 2(\cos \theta +K\sin \theta ):=\widehat{S_{1}^{1}}$, where $%
K=2(c_{1}c_{2}+c_{3}c_{4}+\cdots +c_{s-1}c_{s})$, $0<K\leq 1$.

Using Eq.(\ref{z001}), we obtain 
\begin{eqnarray}
&&\rho _{AB}^{(2,1)}=\frac{1}{2}\rho _{AB}^{(1)}  \nonumber \\
&&+\frac{1}{4}[I_{s}\otimes (I_{\frac{t}{2}}\otimes \sigma _{3})]\rho
_{AB}^{(1)}[I_{s}\otimes (I_{\frac{t}{2}}\otimes \sigma _{3})]  \nonumber \\
&&+\frac{1}{4}[I_{s}\otimes (I_{\frac{t}{2}}\otimes \sigma _{1})]\rho
_{AB}^{(1)}[I_{s}\otimes (I_{\frac{t}{2}}\otimes \sigma _{1})].
\end{eqnarray}%
Then taking $B_{y}^{(2,1)}=B_{y}^{(1,1)}$ for $y=0,1$, we get $S_{2}^{1}\geq
(\cos \theta +K\sin \theta ):=\widehat{S_{1}^{1}}$. The trade-off
relationship between $\widehat{S_{1}^{1}}$ and $\widehat{S_{2}^{1}}$ is
given by 
\begin{equation}
\widehat{S_{2}^{1}}=\frac{1}{2}\widehat{S_{1}^{1}}.  \label{z008}
\end{equation}%
When $\theta =\arctan K$, $\widehat{S_{1}^{1}}\ $attains the maximum value $2%
\sqrt{1+K^{2}}$. At this moment $\widehat{S_{2}^{1}}=\sqrt{1+K^{2}}$.
Moreover, when $|\psi \rangle =\frac{1}{\sqrt{2}}(|00\rangle +|11\rangle )$,
that is, $K=1$, we obtain the same maximum value of $\widehat{S_{1}^{1}}$ as
in Ref.\cite{asat}.

Case (ii): ($\lambda =2$). We take the Bob$_{1}$'s projective measurements
to be 
\begin{equation}
B_{0|0}^{(1,2)}=I_{t}  \label{z004}
\end{equation}%
and 
\begin{equation}
B_{0|1}^{(1,2)}=\frac{1}{2}(I_{t}+(I_{\frac{t}{2}}\otimes \sigma _{1})).
\label{z005}
\end{equation}%
Denote $B_{1|y}^{(1,2)}=I_{t}-B_{0|y}^{(1,2)}$ and $%
B_{y}^{(1,2)}=B_{0|y}^{(1,2)}-B_{1|y}^{(1,2)}$ for $y=0,1$. Similarly, we
can obtain $S_{1}^{2}\geq 2K\sin \theta :=\widehat{S_{1}^{2}}$. By
calculation, we have 
\begin{eqnarray}
&&\rho _{AB}^{(2,2)}=\frac{3}{4}\rho _{AB}^{(1)}  \nonumber \\
&&+\frac{1}{4}[I_{s}\otimes (I_{\frac{t}{2}}\otimes \sigma _{1})]\rho
_{AB}^{(1)}[I_{s}\otimes (I_{\frac{t}{2}}\otimes \sigma _{1})].
\end{eqnarray}%
Then take $B_{0}^{(2,2)}=I_{\frac{t}{2}}\otimes \sigma _{3}$, $%
B_{1}^{(2,2)}=I_{\frac{t}{2}}\otimes \sigma _{1}$, we obtain $S_{2}^{2}\geq
2K\sin \theta +\cos \theta :=\widehat{S_{2}^{2}}$. The trade-off
relationship between $\widehat{S_{1}^{2}}$ and $\widehat{S_{2}^{2}}$ becomes 
\begin{equation}
\widehat{S_{2}^{2}}=\widehat{S_{1}^{2}}+\frac{1}{2K}\sqrt{4K^{2}-(\widehat{%
S_{1}^{2}})^{2}}.  \label{z009}
\end{equation}%
When $\theta =\arctan 2K$, $\widehat{S_{2}^{2}}$ achieves the maximum value $%
\sqrt{4K^{2}+1}$. Meanwhile, $\widehat{S_{1}^{2}}=\frac{4K^{2}}{\sqrt{%
4K^{2}+1}}$. In particular, for $|\psi \rangle =\frac{1}{\sqrt{2}}%
(|00\rangle +|11\rangle )$ ($K=1$), our the trade-off relation Eq. (\ref%
{z009}) gives rise to the Eq.(6) of Ref.\cite{asat}.

Concerning the Bell nolocality, we assume that the probability of choosing
the first (second) measurement is $p$ ($1-p$). According to the definition (%
\ref{z002}), it can be seen that $S_{1}=pS_{1}^{1}+(1-p)S_{1}^{2}\geq p%
\widehat{S_{1}^{1}}+(1-p)\widehat{S_{1}^{2}}=2p\sqrt{1+K^{2}}+(1-p)\frac{%
4K^{2}}{\sqrt{4K^{2}+1}}$ and $S_{2}=pS_{2}^{1}+(1-p)S_{2}^{2}\geq p\widehat{%
S_{2}^{1}}+(1-p)\widehat{S_{2}^{2}}=p\sqrt{1+K^{2}}+(1-p)\sqrt{4K^{2}+1}$.
The nonlocality sharing problem is then transformed to find parameters $p$
and $K$ such that $S_{1}$ and $S_{2}$ are both greater than $2$. As long as
the right-hand formulas of the two inequalities above are both greater than
2 is sufficient. From FIG.2 and FIG.3 we see that they can both be greater
than $2$ for some $p$ and $K$. For example when $K=1$, $p\in \lbrack \frac{2%
\sqrt{5}-4}{2\sqrt{10}-4},\frac{\sqrt{5}-2}{\sqrt{5}-\sqrt{2}}]\approx
\lbrack 0.203,0.286]$, $S_{1}$ and $S_{2}$ are simultaneously greater than
2. Because $S_{1}$ and $S_{2}$ are both continuous functions of $p$ and $K$,
there is still a finite domain in which $S_{1}$ and $S_{2}$ are both greater
than $2$. This also fully demonstrates that projective measurements are
sufficient for sharing Bell nonlocality for bipartite high dimension pure
states. 
\begin{figure}[tbp]
\includegraphics[width=0.45\textwidth]{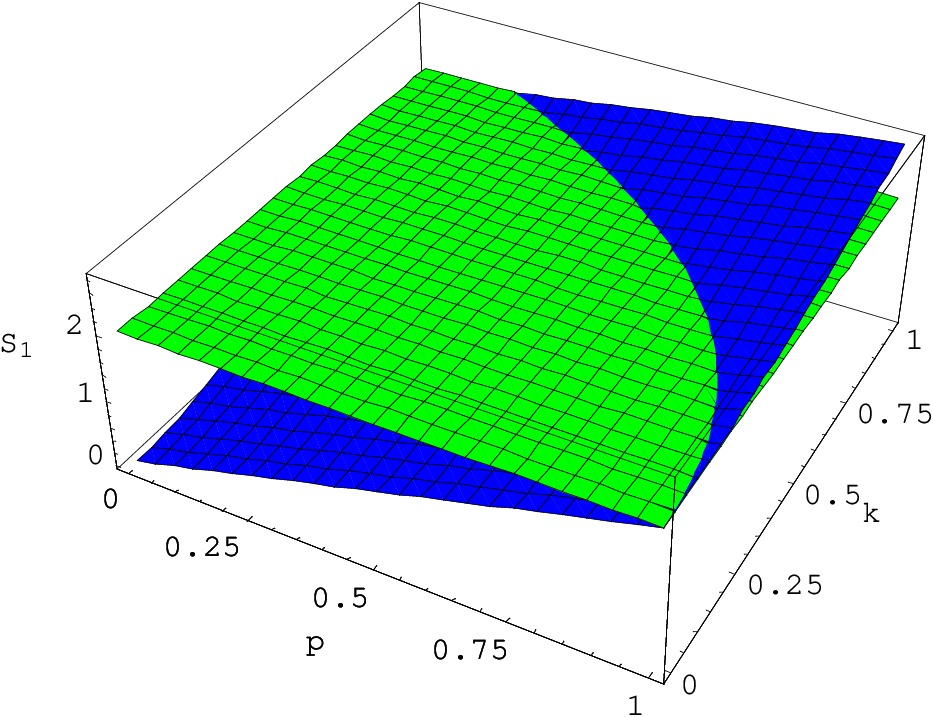}
\caption{$p$ and $K$ as parameters of $S_{1}$ (blue), contour surface $%
S_{1}=2$ (green).}
\end{figure}
\begin{figure}[tbp]
\includegraphics[width=0.5\textwidth]{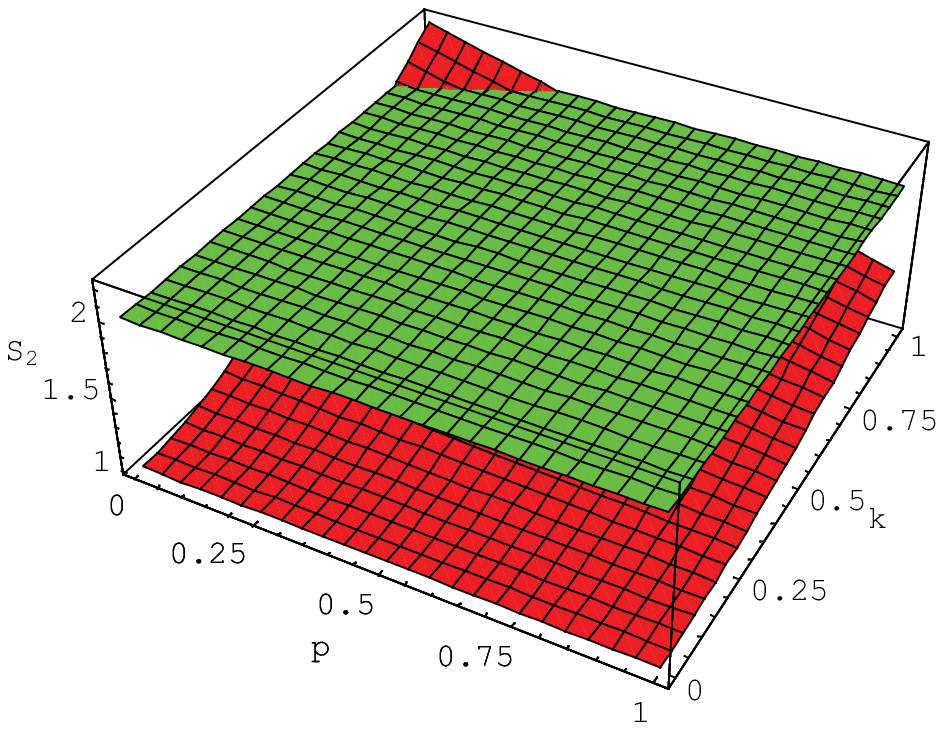}\vspace{-20.0em}
\caption{$p$ and $K$ as parameters of $S_{2}$ (red), contour surface $S_{2}=2
$ (green).}
\end{figure}

When $s$ and $t$ are odd numbers, we only need to take the following
measurement operators and follow the calculation method in Ref.\cite{ztfs}
to obtain the same conclusion as when $s$ is even, except that the
expression of $K$ is changed to be $2(c_1c_2+c_3c_4+\cdots+c_{s-2}c_{s-1})$.
The measurement operators can be selected as 
\begin{equation}
A_{0}=\left(%
\begin{array}{cc}
I_{[\frac{s}{2}]} \otimes(\cos\theta\sigma_3+\sin\theta\sigma_1) & 0 \\ 
0 & 1%
\end{array}%
\right)
\end{equation}
and 
\begin{equation}
A_{1}=\left(%
\begin{array}{cc}
I_{[\frac{s}{2}]} \otimes(\cos\theta\sigma_3-\sin\theta\sigma_1) & 0 \\ 
0 & 1%
\end{array}%
\right)
\end{equation}
for some $\theta\in [0,\frac{\pi}{2}]$, where $[m]$ represents the integer
less or equal to $m$.

Similarly, in case (i) the corresponding projective measurements of Bob$_1$
are taken as 
\[
B^{(1,1)}_{0|0}=\frac{1}{2}[I_t+\left(%
\begin{array}{cc}
(I_{[\frac{t}{2}]}\otimes\sigma_3) & 0 \\ 
0 & 1%
\end{array}%
\right)] ;
\]
\[
B^{(1,1)}_{0|1}=\frac{1}{2}[I_t+\left(%
\begin{array}{cc}
(I_{[\frac{t}{2}]}\otimes\sigma_1) & 0 \\ 
0 & 1%
\end{array}%
\right)]. 
\]
In case (ii), the corresponding projective measurements of Bob$_1$ are taken
as 
\[
B^{(1,2)}_{0|0}=I_t;
\]
\[
B^{(1,2)}_{0|1}=\frac{1}{2}[I_t+\left(%
\begin{array}{cc}
(I_{[\frac{t}{2}]}\otimes\sigma_1) & 0 \\ 
0 & 1%
\end{array}%
\right)]. 
\]
The corresponding projective measurements of Bob$_2$ are taken as 
\[
B_0^{(2,2)}=\left(%
\begin{array}{cc}
(I_{[\frac{t}{2}]}\otimes\sigma_3 & 0 \\ 
0 & 1%
\end{array}%
\right);
\]
\[
B_1^{(2,2)}=\left(%
\begin{array}{cc}
(I_{[\frac{t}{2}]}\otimes\sigma_1 & 0 \\ 
0 & 1%
\end{array}%
\right).
\]
One derives again that the Bell nonlocality of bipartite high dimension pure
states can be shared under projective measurements.

\section{Conclusions and Outlook}

We have shown that projective measurements are sufficient for sharing the
Bell nonlocality of high-dimensional entangled pure states. Namely, two
independent Bobs may share states with a single Alice such that all the
shared states violate the CHSH inequality. Our work greatly expands the
range of quantum states given in Ref.\cite{asat}. These quantum states share
the Bell nonlocality through projection measurements and the distribution of
shared classical randomness. These results are not only theoretically
interesting, but also are of significance for experimental implementation,
since Projective measurements enable one to demonstrate sequential
nonlocality sharing in much simpler setups than previous non-projective
measurements\cite{mlmg,mzxc,glam,tcym,gmmh}. In fact, unsharp quantum
measurements have been proven to be useful for device-independent
self-testing and recycling quantum communication\cite{knab,mjjm,prap}. Our
results show that with only projection measurements it might be feasible for
some sequential quantum information protocols related to quantum coherence%
\cite{sasm}, entanglement witnessing\cite{asau,adaa}, quantum steering\cite%
{sdsa,xhxy} and quantum contextuality\cite{hnrs}.

We have proven that based on POVMs a high-dimension bipartite entangled pure
state may produces $n$ sequential violations of the CHSH inequality\cite%
{ztfs}. For projective measurements it is still unknown about how many
sequence violations can occur for high-dimensional entangled pure states at
most. It is also a meaningful problem to design the optimal projection
measurement scheme. Instead of pure states, one may ask if any special mixed
states can also be used for nonlocality sharing under projection
measurements and shared randomness. Moreover, conclusions about simultaneous
bilateral measurements and shared randomness would be also of importance.

\bigskip Acknowledgments: We thank the anonymous reviewers for their helpful
suggestions. This work is supported by the National Natural Science
Foundation of China (NSFC) under Grant Nos.12204137, 12075159 and 12171044;
the Hainan Provincial Natural Science Foundation of China under Grant
No.121RC539, the specific research fund of the Innovation Platform for
Academicians of Hainan Province under Grant No.YSPTZX202215, Beijing Natural
Science Foundation (Grant No.Z190005).

\end{document}